\documentclass[aps,prl,reprint,groupedaddress,showpacs]{revtex4-1}

\usepackage{amsmath}
\usepackage[pdftex]{graphicx}

\usepackage[utf8]{inputenc}

\usepackage{color} 
\usepackage{hyperref}

\newcommand{\fig}[1]{Fig.~\ref{#1}}
\newcommand{\eq}[1]{Eq.~(\ref{#1})}

\begin{document}

\normalsize

\title{Influence of bonding on superconductivity in high-pressure hydrides}

\author{Christoph Heil}
\email[]{christoph.heil@tugraz.at}
\affiliation{Institute of Theoretical and Computational Physics, Graz University of Technology, NAWI Graz, 8010 Graz, Austria}

\author{Lilia Boeri}
\affiliation{Institute of Theoretical and Computational Physics, Graz University of Technology, NAWI Graz, 8010 Graz, Austria}

\date{\today}

\begin{abstract}
The recent reports on high-temperature superconductivity above 190\;K in hydrogen sulfide at 200\;GPa pressure, exceeding all previously discovered superconductors, has greatly invigorated interest in dense hydrogen-rich solids. Here, we investigate a possible way to optimize the critical temperature in these compounds by using first-principles linear-response calculations. We construct hypothetical {\em alchemical} atoms to smoothly interpolate between elements of the chalcogen group and study their bonding and superconducting properties. Our results show that the already remarkable critical temperatures of H$_3$S could be improved even further by increasing the ionic character of the relevant bonds, i.e., partially replacing sulfur with more electronegative elements.
\end{abstract}


\maketitle

The recent experimental report of superconductivity with a critical temperature ($T_c$) exceeding $190$\;K in dense hydrogen 
sulfide~\cite{drozdov_conventional_2014} could be one of the most important discoveries in the field of superconductivity, if confirmed. Although x-ray scattering data are lacking due to the extreme pressures involved ($p \sim 200$\;GPa), there is a widespread consensus that the superconducting samples are composed of a sulfur-trihydride (H$_3$S) bcc phase predicted by Duan {\em et al.}~\cite{duan_pressure-induced_2014}. Theoretical studies~\cite{li_metallization_2014,papaconstantopoulos_cubic_2015, flores-livas_high_2015,errea_high-pressure_2015,akashi_fully_2015,bernstein_what_2015,nicol_comparison_2015} and isotope effect measurements~\cite{drozdov_conventional_2015} unambiguously indicate that, unlike the high-$T_c$ iron pnictides and cuprates, H$_3$S is a conventional BCS~\cite{bardeen_theory_1957} superconductor.

Phonon-mediated superconductors are accurately described by the Migdal-Eliashberg theory~\cite{migdal_interaction_1958,eliashberg_interactions_1960}; their $T_c$ is well approximated by the Allen-Dynes formula~\cite{allen_transition_1975},
\begin{equation}
 T_c = \frac{\omega_\text{ln}}{1.2} \exp  \left( \frac{-1.04 \, (1+\lambda)}{\lambda-\mu^*(1+0.62 \lambda)} \right)~, \label{eq:Tc}
\end{equation}
where $\lambda$ is the electron-phonon ($e$-$p$) coupling constant, $\omega_\text{ln}$ is the logarithmic average of phonon frequencies, and $\mu^*$ is the Coulomb pseudopotential. Thus, Migdal-Eliashberg theory indicates a clear path to optimize $T_c$, i.e., looking for materials with high characteristic phonon frequencies ($\omega_\text{ln}$) and strong $e$-$p$ coupling constants $\lambda$.

In a seminal paper that is now almost 50 years old, Ashcroft noticed that these conditions could be realized in ultra-dense metallic hydrogen~\cite{ashcroft_metallic_1968}, yet only the recent progress in experimental techniques for high-pressure research and improved methods for crystal-structure prediction have made this hypothesis viable~\cite{cudazzo_textitabinitio_2008}. In the last few years, there has been significant interest in the superconducting behavior of hydrogen-rich solids at high pressures~\cite{ashcroft_hydrogen_2004,sun_optical_2006,feng_structures_2006,tse_novel_2007,eremets_superconductivity_2008,gao_superconducting_2008,zurek_little_2009,errea_first-principles_2013}. These compounds display a fascinating variety of crystal structures, as some contain molecular hydrogen, in others the H-H bond is broken, and hydrogen forms direct bonds with the other atoms. The predicted $T_c$'s close to ambient pressures are just a few K, while for pressures above  $150$\;GPa they range from 
$64$ to $235$\;K.

Among these compounds, H$_3$S has been predicted to exhibit one of the highest $T_c$'s. Reference~\cite{bernstein_what_2015} underlined that the mechanism boosting its $T_c$ is essentially the same one that occurs in magnesium diboride (MgB$_2$), which broke the record of critical temperatures for conventional superconductors in 2001~\cite{an_superconductivity_2001,lee_superconductivity_2004,boeri_three-dimensional_2004}. In MgB$_2$, the presence of electronic bands with a strong covalent character at the Fermi level leads to $e$-$p$ matrix elements much larger than those found in other conventional superconductors. The experience with this material has shown that, besides the well-known strategy of carrier doping, a very promising route to increase $T_c$ is to act on the bonding properties of the materials, directly influencing the $e$-$p$ matrix elements. Within the range of MgB$_2$-like materials, the highest critical temperature was predicted for hole-doped LiBC, whose stiff ionic bonds are even more 
favorable for 
superconductivity~\cite{rosner_prediction_2002}.

H$_3$S is very similar to MgB$_2$ as it also exhibits strong covalent bonds giving rise to large electron-phonon couplings~\cite{papaconstantopoulos_cubic_2015,bernstein_what_2015,errea_high-pressure_2015,flores-livas_high_2015,akashi_fully_2015} and strong anharmonic effects~\cite{errea_high-pressure_2015,yildirim_giant_2001,boeri_small_2002,lazzeri_anharmonic_2003}, and although its critical temperature is already spectacularly high, there is no reason to expect that it cannot be increased any further.

The aim of this Rapid Communication is to explore whether the superconducting properties of H$_3$S can be further enhanced by increasing (or decreasing) the covalent character of the sulfur-hydrogen bond. We perform a systematic study of the electron-phonon properties of several H$_3X$ compounds in the bcc high-pressure structure of H$_3$S, using linear-response calculations~\footnote{All calculations have been performed with the \textsc{Abinit} linear-response DFT package~\cite{abinit_2015}. We used Troullier-Martins pseudopotentials~\cite{troullier_efficient_1991} within the generalized gradient approximation~\cite{perdew_generalized_1996} and a kinetic energy cut-off for the planewaves of $85$\;Ry. The groundstate wavefunction and the charge density have been calculated using a $32^3$ Monkhorst-Pack $\mathbf{k}$-point grid and a tetrahedron method. Phonon properties were calculated on a $8^3$ Monkhorst-Pack $\mathbf{q}$-point grid and interpolated with a real space Fourier procedure. We counterchecked our 
results, whenever 
possible, with Wien2k and \textsc{Quantum Espresso}.}. The $X$ atom is either one of the first four atoms of the chalcogen group (O, S, Se, Te), or a hypothetical {\em alchemical} mixture of neighboring chalcogens (i.e., O and S, S and Se, and so on)~\footnote{The {\em alchemical} mixing of pseudopotentials is equivalent to the virtual crystal approximation if the two atoms are neighbors in the periodic table. \mbox{$V_\text{VCA} = \gamma V_{X_1} + (1-\gamma) V_{X_2}~, \; \gamma \in [0,1]$}, and analogously for the mass.}\footnote{Previous studies on the thermodynamic stability in chalcogen hydrides exist, see Refs.~\cite{duan_pressure-induced_2014,flores-livas_high_2015,errea_high-pressure_2015,bernstein_what_2015,zhang_phase_2015,zhong_tellurium_2015}}. This approximation allows us to study the influence of covalency and atomic size on the bonding, dynamical stability, and electron-phonon coupling of the H$_3X$ structure in a controlled way, without introducing spurious effects related to doping. We will 
also 
introduce quantities to measure the bonding characteristics of dense hydrogen-rich compounds that will be useful in the search for new high-$T_c$ materials with high-throughput methods~\cite{curtarolo_high-throughput_2013}. In order to minimize the number of external parameters, we work at a fixed pressure~\footnote{The lattice constant given by Ref.~\cite{duan_pressure-induced_2014} results in a pressure of $190$\;GPa in our calculations, which we have kept constant for all compounds.}.

\begin{figure}[t]
  \begin{center}
    \includegraphics[width=1.0\linewidth]{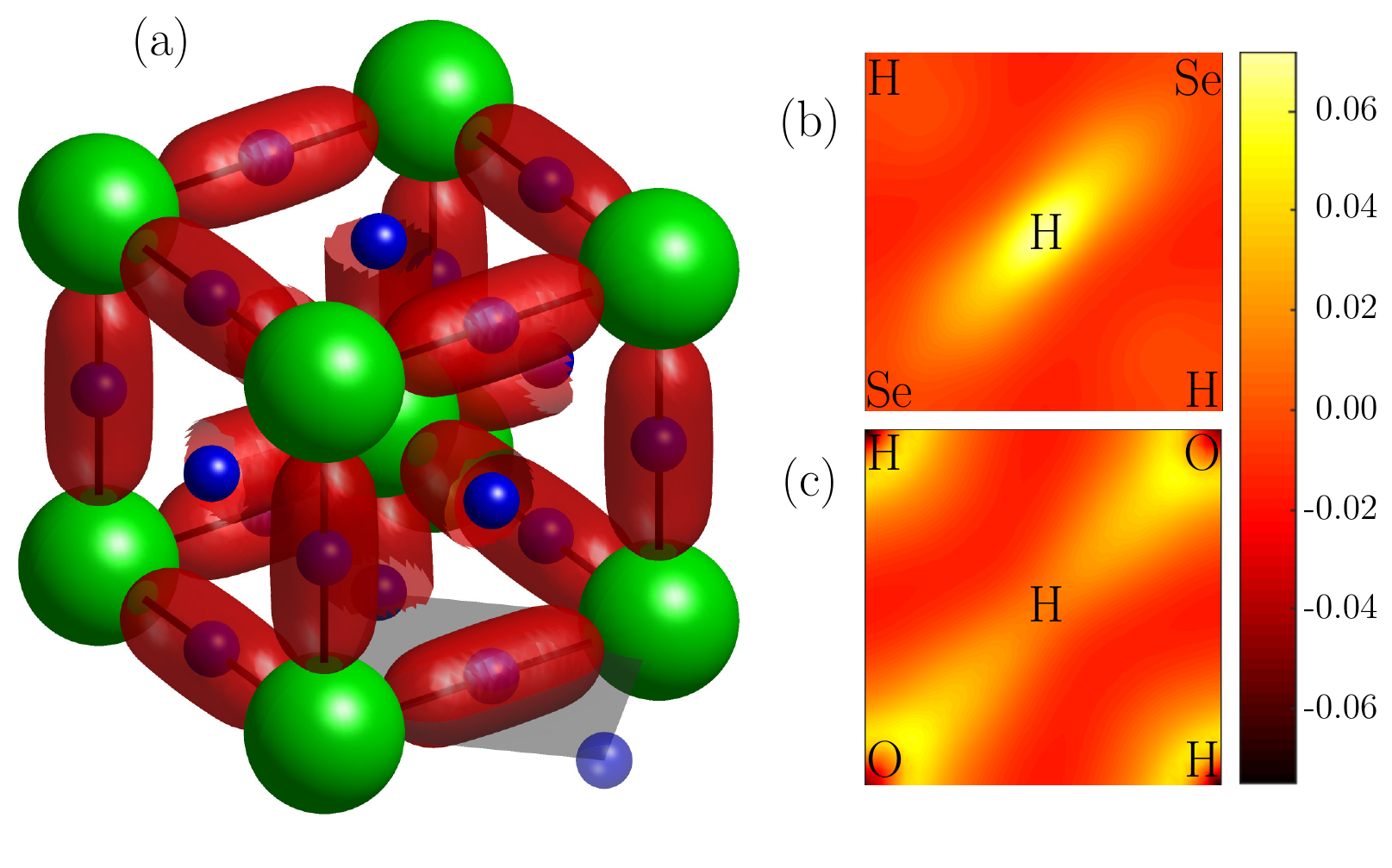}
    \vspace*{-2em}
  \caption{(Color online) (a) Crystal structure and isocharge surface ($\rho=0.01$) of H$_3$Se in the conventional unit cell. The Se atoms are indicated by green spheres, and the hydrogen atoms by blue spheres. The red surfaces are isocharge surfaces (see text). (b) Charge plot in the gray-shaded plane of (a) for H$_3$Se and (c) for H$_3$O, both with substracted backgrounds.}
  \label{fig:substracted_charge}
  \end{center}
\end{figure}

Figure~\ref{fig:substracted_charge} shows the $Im \bar{3} m$ bcc structure of the H$_3X$ compounds at high pressures: The $X$ atoms (green) occupy the $2a$ Wyckoff position and form a bcc lattice, while the hydrogen atoms (H, blue) occupy the $6b$ Wyckoff positions in between. Each H forms two short and four long bonds with its $X$ neighbors in the ($100$), ($110$) and ($1\bar{1}0$) directions. When the electronegativity of the $X$ atom is slightly larger than that of H~($2.1$), as in S~($2.4$) and Se~($2.5$), the charge accumulates around the center of the short $X$-H-$X$ bonds, giving them a strongly {\em covalent} character. The red, H-centered ellipsoids in \fig{fig:substracted_charge}(a) delimit regions in which the charge density of H$_3$Se is higher than that of hypothetical Se and H$_3$ solids in the same crystal structure. A contour plot of the same quantity in the basal plane is plotted in \fig{fig:substracted_charge}(b). As the $X$-H-$X$ bonds acquire an increasing ionic character by increasing 
the 
electronegativity of $X$, the maximum of the charge density distribution moves from the vicinity of hydrogen to that of the $X$ atom; this is seen in the contour plot for H$_3$O, shown in \fig{fig:substracted_charge}(c).

\begin{figure}[t]
    \hspace*{-1em} \includegraphics[width=1.06\linewidth]{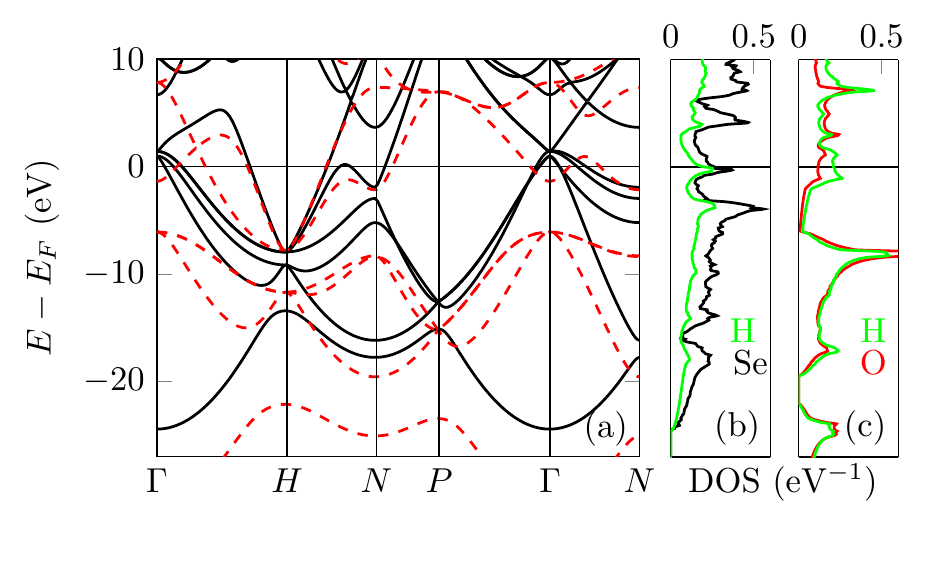}
    \vspace*{-3em}
  \caption{(Color online) (a) Electronic band structure of H$_3$Se (solid black) and H$_3$O (dashed red) along the path $\Gamma$(0,0,0)--$H$(0,1,0)--$N$(0.5,0.5,0)--$P$(0.5,0.5,0.5)--$\Gamma$(0,0,0)--$N$(0.5,0.5,0) in units of $2\pi/a$. (b) Partial DOS of H$_3$Se with Se (one H) contributions in black (green). (c) Partial DOS of H$_3$O with O (one H) contributions in red (green).}
  \label{fig:electrons}
\end{figure}

We will now discuss in detail how and why the electronic and vibrational properties of these two compounds differ. We start from \fig{fig:electrons}(a), where we show the electronic structure of the two compounds. The band structure of H$_3$Se (solid black) is  in nice agreement with previous calculations and, as discussed by several authors, very similar to the one of H$_3$S~\cite{bernstein_what_2015,errea_high-pressure_2015,flores-livas_high_2015}. Se- and H-derived electronic states are strongly hybridized, especially in the vicinity of the Fermi energy. This leads to several avoided crossings and eventually to a strong van-Hove singularity slightly below $E_F$~\footnote{We also performed rigid-band calculations (not shown) for H$_3$S to investigate charge doping effects. These calculations, however, revealed that $\lambda$ and $T_c$ cannot be significantly increased with charge doping}.

At the same pressure, H$_3$O would have a lattice constant $\sim$\;$25$\% smaller than H$_3$Se; its band structure is plotted as dashed red lines in \fig{fig:electrons}(a). The remarkable difference in the electronic structure of the two compounds is due to the large electronegativity of O, which increases the hybridization between O and H states and leads to even larger avoided crossings. As a consequence, the van Hove singularity close to $E_F$ splits in two peaks, which move apart as well.

\begin{figure}[t]
    \hspace*{-1em}  \includegraphics[width=1.06\linewidth]{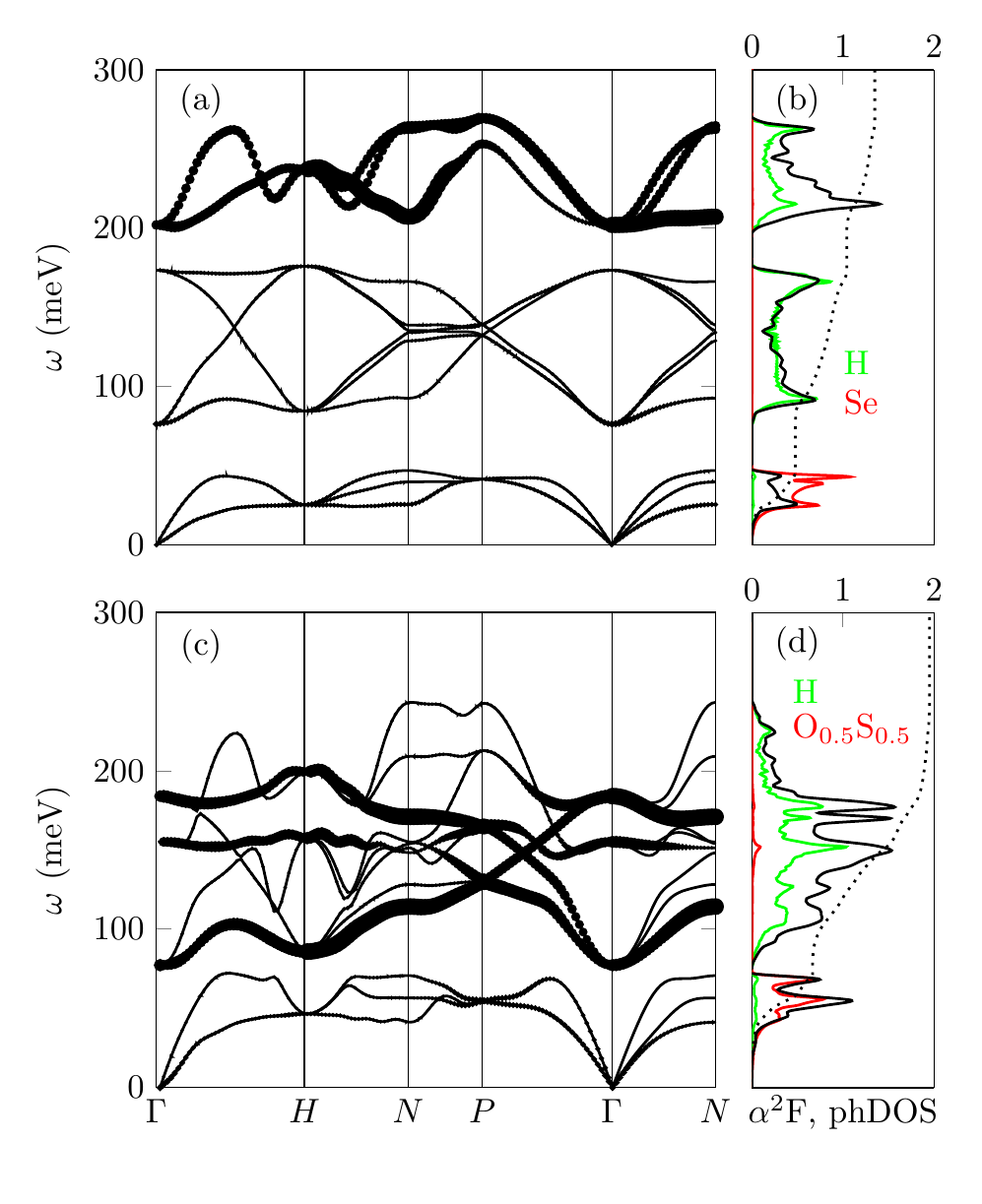}
    \vspace*{-3em}
  \caption{(Color online) Phononic band structure of (a)~H$_3$Se and (c)~H$_3$O$_{0.5}$S$_{0.5}$, where the size of the circles indicates the amount of bond-stretching character. $\alpha^2 F(\omega)$ (black), integrated $\lambda$ (black dotted), and phonon DOS (Se character red, H character green) of (b)~SeH$_3$ and (d)~H$_3$O.}
  \label{fig:phononic}
\end{figure}

Figure~\ref{fig:phononic} shows the corresponding vibrational properties: Phonon dispersions are plotted in Figs.~\ref{fig:phononic}(a) and \ref{fig:phononic}(c), while Figs.~\ref{fig:phononic}(b) and \ref{fig:phononic}(d) show the partial phonon density of states (DOS) and the $e$-$p$ spectral function  $\alpha^2 F(\omega)$,
\begin{equation}
 \alpha^2 F(\omega) = \frac{1}{N_{E_F}} \sum \limits_{\mathbf{k} \mathbf{q},\nu} |g_{\mathbf{k},\mathbf{k}+\mathbf{q},\nu}|^2 \delta(\epsilon_\mathbf{k}) \delta(\epsilon_{\mathbf{k}+\mathbf{q}}) \delta(\omega-\omega_{\mathbf{q},\nu}) \label{eq:a2F}
\end{equation}
where $N_{E_F}$ is the DOS at the Fermi level, $\omega_{\mathbf{q},\nu}$ is the phonon frequency of mode $\nu$ at wavevector $\mathbf{q}$, and 
$|g_{\mathbf{k},\mathbf{k}+\mathbf{q},\nu}|$ is the electron-phonon matrix element~\cite{allen_neutron_1972}, between two electronic states with momenta $\mathbf{k}$ and $\mathbf{k+q}$ at the Fermi level. The dotted line shows the $\omega$-dependent $e$-$p$ coupling parameter $\lambda (\omega) = 2 \int_0^{\omega} d \Omega \frac{\alpha^2 F(\Omega)}{\Omega}$; $\lambda (\infty)$ gives the total $e$-$p$ coupling parameter already discussed in \eq{eq:Tc}.

In H$_3$Se [Fig.~\ref{fig:phononic}(a)] the acoustic modes extend up to an energy of about $50$\;meV. The first six optical branches are separated by a gap of around $25$\;meV from the acoustic ones and have a bandwidth of $\sim$$100$\;meV. Separated again by an energy gap of $25$\;meV are the last three optical branches. These three modes have a mainly bond-stretching character, according to the definition of Ref.~\cite{errea_high-pressure_2015}. In turn, the middle six phonon branches have a mostly bond-bending oscillatory character. The acoustic modes contribute \mbox{$\sim 1/3$} to the total value of $\lambda$. The $\alpha^2 F(\omega)$ of the bond-stretching modes is distinctly different from the shape of the phonon DOS and adds about $1/4$ to the total $\lambda$. For H$_3$Se, we obtain $\omega_\text{ln}=84$\;meV and $\lambda=1.35$, leading to a $T_c=100$\;K for $\mu^* = 0.1$.

The superconducting properties improve for the smaller-$Z$ chalcogens, which have smaller atomic masses, decreasing ionic radii, and increasing electronegativities. In agreement with previous calculations, we find that the $T_c$ of H$_3$S is distinctly larger than that of H$_3$Se (for H$_3$S we get \mbox{$\omega_\text{ln}=97$\;meV,} \mbox{$\lambda=1.83$}, and \mbox{$T_c=151$\;K} for \mbox{$\mu^* = 0.1$})~\footnote{We want to remark that the $\lambda$ values we obtain are systematically lower than in other works, which use different codes. Ref.~\cite{flores-livas_high_2015} for example computed $\lambda=2.41$ and $\omega_\text{ln}=109$\;meV for H$_3$S, which are substantially larger than ours. Apart from numerical differences due to the different method for $\mathbf{k}$-space integration, there is a significant difference in the dispersion of the lower bond-stretching modes, which in our calculations are much harder than in Ref.~\cite{flores-livas_high_2015}. Errea {\em et al.}~\cite{errea_high-pressure_
2015} point out a strong phonon anharmonicity in H$_3$S, thus even small differences in the computational details used for linear response calculations may result in discernible deviations in the estimates obtained with the harmonic approximation. In fact, our phonon dispersion, $\lambda$ and $\omega_\text{ln}$ values are much closer to the values reported in Ref.~\cite{errea_high-pressure_2015}.}. In addition, we find that a further improvement of the superconducting properties occurs in the oxygen-rich part of our {\em alchemical} phase diagram, where the $X$-H-$X$ bond becomes more and more ionic, before the dynamical stability of the lattice breaks down eventually for oxygen contents larger than $50$\%. Indeed, H$_3$O should crystallize in the $Im\bar{3}m$ structure only above $11000$\;GPa~\cite{benoit_new_1996,zhang_h$_mathbf4$o_2013}.

Figure~\ref{fig:phononic}(c) shows the electron-phonon properties of the last stable compound, i.e., H$_3$O$_{0.5}$S$_{0.5}$. The acoustic branches 
extend up to $\sim$$70$\;meV and contribute again about $1/3$ to the total $\lambda=1.95$. The optical modes are located in the energy window from $80$ to $245$\;meV. Bond-stretching modes are mainly responsible for the peaks in $\alpha^2 F(\omega)$ centered at $110$, $150$, and $175$\;meV. Integrating $\alpha^2 F(\omega)$ for H$_3$O$_{0.5}$S$_{0.5}$, we obtain $\omega_\text{ln}=100$\;meV and $T_c=164$\;K for $\mu^*=0.1$.

\begin{figure}[htbp]
    \hspace*{-1em} \includegraphics[width=1.06\linewidth]{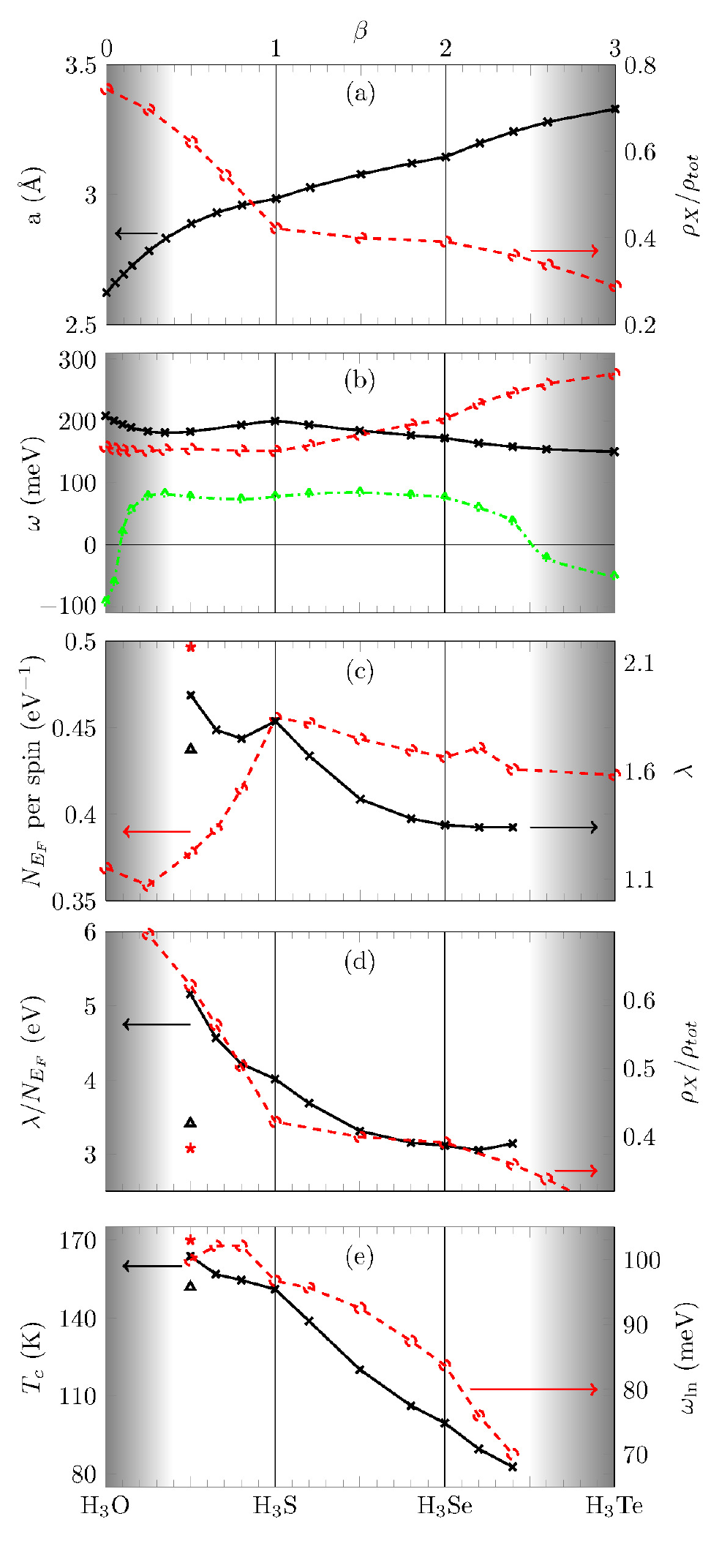}
    \vspace*{-3em}
  \caption{(Color online) (a) Lattice constant $a$ (solid black) and partial charge $\rho_X/\rho_\text{tot}$ (dashed red). (b) Optical phonon frequencies at $\Gamma$. (c) $\lambda$ (solid black) and $N_{E_F}$ (dashed red). (d) $\lambda/N_{E_F}$ (solid black) and rescaled partial charge $\rho_X/\rho_\text{tot}$ (dashed red). (e) $T_c$ (solid black) and $\omega_\text{ln}$ (dashed red). The isolated symbols in (c)-(e) indicate results for H$_3$S calculated at the same lattice constant as H$_3$O$_{0.5}$S$_{0.5}$ (see text).}
  \label{fig:overview}
\end{figure}

We have so far discussed only specific choices of $X$. To comprehend how individual differences in bonding properties, dynamical stability, and electron-phonon coupling translate into actual trends in an {\em alchemical} phase diagram, we present in \fig{fig:overview} an overview of representative quantities as a function of a continuous variable $\beta$ that ranges from zero (oxygen) to three (tellurium); non-integer values represent {\em alchemical} pseudoatoms.

Figure~\ref{fig:overview}(a) shows in black the optimized lattice constant, which increases smoothly with $\beta$ from $a=2.62$ to $3.33$\;\r{A}. In red, we plot a compact measure of the charge localization along the short bond direction, i.e., the fraction $\rho_X/\rho_\text{tot}$ of total charge which is localized around the $X$ atom~\footnote{What is computed in practice is the percentage of charge density contained outside the volume in which 90$\%$ of the total charge of a hypothetical H$_3$ solid in the same unit would be found.}. This quantity increases as $\beta \to 0$, as the maximum of the charge distribution shifts smoothly from the H to the $X$ atom. In \fig{fig:overview}(b) we show the frequency evolution of the triply degenerate optical phonon branches at the $\Gamma$ point -- see \fig{fig:phononic}. All branches have a (partially) H bond-stretching and bond-bending character and evolve smoothly with $\beta$. While the green and red branches have $T_{1u}$ symmetry, the black one is of $T_{2u}$ 
symmetry. For high O and Te contents, the lowest branch displays imaginary frequencies, indicating a dynamical instability of the structure~\cite{benoit_new_1996,zhong_tellurium_2015}. In O-rich compounds, another instability occurs earlier near the $P$ point. The gray-shaded areas in the figure indicate parts of the $\beta$ phase diagram where the H$_3X$ structure is dynamically unstable.

In \fig{fig:overview}(c), we plot in red the value of the electronic DOS at the Fermi level $N_{E_F}$; it is essentially constant for S$\rightarrow$Se$\rightarrow$Te ($\beta > 1$), and decreases sharply for $\beta \to 0$, due to the increased O-H hybridization -- see \fig{fig:electrons}.   
On the other hand, $\lambda$ is almost monotonously increasing for decreasing $\beta$; $\omega_\text{ln}$ [\fig{fig:overview}(e)] exhibits a similar behavior, i.e., it remains nearly constant between O and S, and bends downwards for Se and Te. 
A large value of $\omega_\text{ln}$ for small $\beta$ is not in contradiction with the incipient lattice instability, since, as mentioned earlier, the phonon modes which undergo softening carry only a small fraction of the total $e$-$p$ coupling~\cite{lee_superconductivity_2004,boeri_three-dimensional_2004,yao_superconductivity_2009}. The combined effect of $\omega_\text{ln}$ and $\lambda$ thus results in a net monotonous increase of $T_c$, from $83$\;K on the Te-rich side to $164$\;K on the O-rich side, as shown as the red dashed line in \fig{fig:overview}(e)~\footnote{As $\beta$ further approaches the transition point, $\omega_\text{ln}$ is expected to decrease rapidly, limiting the maximum $T_c$ achievable.}.

Figure~\ref{fig:overview}(c) also shows that the behavior of $\lambda$ does not correlate with $N_{E_F}$, especially for $\beta \to 0$. Instead, the average $e$-$p$ coupling matrix element $\lambda/N_{E_F}$, plotted in \fig{fig:overview}(d), increases by $\sim 50$\% when going from H$_3$Se to H$_3$O. To rule out that this increase is not just a lattice-constant (pressure) effect, we additionally performed \mbox{$e$-$p$} calculations for H$_3$S at the same lattice constant as H$_3$O$_{0.5}$S$_{0.5}$, i.e., at a pressure of $265$\;GPa: The results, shown as isolated triangles and stars in Figs.~\ref{fig:overview}(c) and \ref{fig:overview}(d) are in line with those of the other chalcogen hydrides ($\lambda=1.70$, $\omega_\text{ln}=103$\;meV, and $T_c=152$\;K). The behavior of $\lambda/N_{E_F}$ strongly resembles that of $\rho_X/\rho_{tot}$, plotted in the same panel as the red lines. This is a clear indication that increasing the ionic character of the $X$-H-$X$ bond can be beneficial for superconductivity. 
Although the increase in $T_c$ is not as 
spectacular as in LiBC compared to MgB$_2$~\cite{rosner_prediction_2002}, our findings indicate increased ionicity as a clear path that experimentalists could follow to optimize $T_c$ in high-pressure hydrides.

In summary, we have studied a possible route to optimize the critical temperature in recently discovered high-$T_c$ superconducting hydrides using first-principles linear-response calculations. We constructed hypothetical {\em alchemical} atoms, which smoothly interpolate atoms of the chalcogen group, and studied their phonon and electron-phonon coupling properties. The effect of {\em alchemical} doping is qualitatively very different from charge doping or pressure, as it allows one to tune the bonding properties of a material, crucially affecting the $e$-$p$ matrix elements. We observed the existence of a ``sweet spot'' for electron-phonon superconductivity in H$_3$O$_{0.5}$S$_{0.5}$, for which the bond is ionic enough to profit from increased electron-phonon coupling, but not so strongly as to lead to a structural instability. 

This implies that, although H$_3$S is already almost optimal for electron-phonon coupling superconductivity, its critical temperature
could be improved further making the $X$-H-$X$ bond more ionic, i.e.,  partially replacing sulfur with more electronegative atoms, such as oxygen or members of the halogen group. Together with the compact measure we provided to characterize the bonding situation, our results highlight promising paths in the search for new conventional high-$T_c$ superconductors with high-throughput methods~\cite{curtarolo_high-throughput_2013}.

{\em Note added.} Recently, we became aware of a similar investigation~\cite{ge_possible_2015}.

We acknowledge fruitful discussions with Gianni Profeta at the beginning of this work. 
This research was supported by the Deutsche Forschungsgemeinschaft under Priority Program No.~1458, 
Grant No.~Boe/3536-1, and from the FWF under the SFB ViCoM F41 P15. Calculations have been performed on the 
dCluster of the Graz University of Technology and the VSC3 of the Vienna University of Technology.

\end{document}